\documentclass{wqcd03}                 

\usepackage{txfonts}                   
\newcommand{\be}[0]{\begin{equation}}
\newcommand{\ee}[0]{\end{equation}}
\newcommand{\ba}[0]{\begin{eqnarray}}
\newcommand{\ea}[0]{\end{eqnarray}}

\confname{QCD@Work 2003 - International Workshop on QCD,
Conversano, Italy, 14--18 June 2003}

\title{Polarized parton distribution functions in
 the valon model framework, using QCD fits to Bernstein polynomials}
\author{Ali N. Khorramian\addressmark{a,b}\thanks{Speaker at the Workshop.}\thanks{khorramiana@theory.ipm.ac.ir}, A. Mirjalili\addressmark{b,c} and
 S. Atashbar Tehrani \addressmark{b,d} }

\address[a]{Physics Department, Semnan University, Semnan, Iran }
\address[b]{Institute for Studies in Theoretical Physics and Mathematics (IPM), Tehran, Iran}
\address[c]{Physics Department, Yazd University, Yazd, Iran}
\address[d]{Physics Department, Persian Gulf University, Boushehr, Iran}

\begin{document}

\begin{abstract}
In this paper polarized valon distribution is derived from
unpolarized valon distribution. In driving polarized valon
distribution some unknown parameters exist which must be
determined by fitting to experimental data. Here we have used Bernstein
polynomial method to fit QCD predictions for the moments
of $g_1^p$ structure function, to suitably the constructed appropriate average quantities of the E143 and SMC experimental data. After calculating
polarized valon distributions and all parton distributions in a
valon, polarized parton density in a proton are available. The
results are used to evaluate the spin components of proton. It
turns out that the results of polarized structure function are in
good agreement with all available experimental data on $g_{1}^p$
of proton.

\end{abstract}
\maketitle
\section{INTRODUCTION}
In valon model, the hadron
is envisaged as a bound state of valence quark cluster called
valon. For example, the bound state of proton consists
of two $U$ and one $D$ valons. These valons thus bear the
quantum numbers of respective valence quarks. Hwa \cite{1} found
the evidence of valons in the deep inelastic neutrino scattering
data , suggested their existence and applied it to a variety of
phenomena.  Hwa and Yang published two
papers \cite{1} and improved the idea of valon model and extracted
new results for the valon distributions. Recently we published a
paper \cite{2} in which we extracted unpolarized constituent quark
and hadronic structure function in Next-to-Leading order. We were
interested here to extend valon model to polarized ones which has not yet been done.\\
\section{Unpolarized and Polarized Valon Distributions}
Valence  and its associated sea quarks plus gluons in the dressing
processes of QCD is defined as {\it{valon}}. In a bound state
problem these processes are virtual and a good approximation for
the problem is to consider a valon as an integral unit whose
internal structure cannot be resolved. The proton, for example, has three valons which, on
the one hand, interacts with each other in a way that is
characterized by the valon wave function and which on the other
hand responds independently in an inclusive hard collision with a
$Q^{2}$ dependence that can be calculated in QCD at high $Q^{2}$.
This picture suggests that the structure function of a hadron
involves a convolution of two distributions: valon distribution in
proton and parton distributions in a valon. In an unpolarized
situation we may write:
\begin{equation}
F_{2}^{p}(x,Q^{2})=\sum_{v}\int_{x}^{1} dy G_{v/p}(y)
{\cal{F}}_{2}^{v}(\frac{x}{y},Q^{2})\;,
\end{equation}
the summation is over the three valons. Here $F_{2}^{p}(x,Q^{2})$
is a proton structure function and ${\cal{F}}_{2}^{v}$ is the
corresponding structure function of a $v$ valon and $G_{v/p}(y)$
indicates the probability for the $v$ valon to have momentum
fraction $y$ in the proton. In Ref. [1] the
unpolarized valon distribution in proton with a new set of
parameters for $G_{U/p}(y)$ and $G_{D/p}(y)$ has been recalculated.
Now we can construct the polarized valon
distributions in proton which are based on the definitions of
unpolarized valon distributions. The polarized valon is defined to
be a dressed polarized valence quark in QCD with the cloud of
polarized gluons and sea quarks which can be resolved by high
$Q^{2}$ probes. In \cite{3} we found the following constrain for
the polarized parton distributions, $|\delta f(x,Q^2)|$, and
unpolarized one, $f(x,Q^2)$, at low value of $Q^2$ which by
positivity requirements implying
\[
|\delta f(x,Q^2)|\leq f(x,Q^2)\;,
\]
where $f=u,\bar u, d, \bar d, s, \bar s, g$, and furthermore by
the sum rules
\[
\Delta u+\Delta \bar u-\Delta d+\Delta \bar
d=A_3=1.2573\pm0.0028\;,
\]
\[
\Delta u+\Delta \bar u+\Delta d+\Delta \bar d-2 (\Delta s+\Delta
\bar s)=A_8=0.579\pm0.025\;,
\]
here $\Delta f$ is the first ($n=1$) moment of $\delta f$, which is defined by
\begin{equation}
\Delta f(Q^2)=\int_{0}^{1}dx \delta f(x,Q^2)\;.
\end{equation}
 For determination polarized parton distributions, the
fundamental analysis is to relate the polarized input density to
unpolarized ones using some intuitive theoretical argument as
guide lines. So we can introduce the following equations to relate
the Non-singlet(NS) and Singlet(S) polarized valons to unpolarized
valon distributions
\[
\delta G_p^{NS} (y)\equiv 2 \delta G_{U/p}(y)+\delta G_{D/p}(y)
\]
\begin{equation}
\;\;\;\;\;\;\;=2 \delta F _{U}(y)\times G_{U/p}(y) + \delta F _{D}(y)\times G_{D/p}(y)\;,
\end{equation}
\[
\delta G_p^{S} (y)\equiv2 f(y)\;\delta G_{U/p}(y) + f(y)\;\delta G_{D/p}(y)
\]
\begin{equation}
\;\;\;\;\;\;\;=2 f(y) \;\delta F _{U}(y)\times G_{U/p}(y) + f(y) \;\delta F _{D}(y)\times G_{D/p}(y)\;,
\end{equation}
where $\delta F _{j}(y)$ and $f(y)$ are defined by:
\begin{equation}
\delta F _{j}(y)=N_{j}y^{\alpha _{j}}(1-y)^{\beta _{j}}(1+\gamma
_{j}y+\eta _{j}y^{0.5})\;,
\end{equation}
\begin{equation}
f(y)=\kappa y^{0.5}+\lambda y+\mu y^{1.5}+\nu y^{2}+\rho y^{2.5}+\tau y^{3}.
\end{equation}
the subscript $j$ refer to $U,D$, and $G_{U/p}(y)$, $G_{D/p}(y)$
has been defined in Ref.[1]. The factor 2 in Eqs.(3,4) backs to existence of 2-$U$ type valons. By using experimental data for $g_1^p$ \cite{4} and using
Bernstein polynomials which we will explain them in Sec. 4, we did a
fitting, and could get parameters of Eqs. (5,6) which are defined
by unpolarized valon distributions $U$ and $D$ in Eqs. (3,4).
According to the Eq. (2), we can write as an example, the
first moment of polarized $u$-valence quark in valon model as
following:
\[
\Delta u_v=\int_{0}^{1}dx\delta u_v=2\int_{0}^{1}dy \delta
G_{U/p}(y),
\]
therefore we can construct for polarized valon distribution the
following sum rules
\[
2\int_{0}^{1}dy \delta G_{U/p}(y)-\int_{0}^{1}dy \delta
G_{D/p}(y)=A_3\;,
\]
\[
2\int_{0}^{1}dy \delta G_{U/p}(y)+\int_{0}^{1}dy \delta
G_{D/p}(y)=A_8\;.
\]
Here we have not considered the SU(3) symmetry breaking where with
totally flavor-symmetric  we have for light sea density $\delta \bar u=\delta
\bar d=\delta \bar s$. So what we will get for $A_3$ and $A_8$
values will have a little bit different with respect to the quoted
values in the above.
\section{Moment of $g_1^p$ polarized structure function}
Let us define the Mellin moments of any structure function
$h(x,Q^2)$ as following: \be
{{\cal{M}}}(n,Q^2)\equiv\int_{0}^{1}x^{n-1} h(x,Q^2) \,dx\;. \ee
Correspondingly in $n$-moment space we indicate the moments of
polarized valon distributions for NS and S sector, $\Delta
G_p^{NS}(n)$ and $\Delta G_p^{S}(n)$ as: \be \Delta G_p^{NS,S}(n)=
\int_{0}^{1}y^{n-1}\delta G_p^{NS,S} (y){dy}\;, \ee
 For the moments of polarized singlet and non singlet distributions we
shall use, the leading order solutions of the renormalization
group equation in QCD. They can be expressed entirely in terms of
the evolution parameter $s$
\begin{equation}
s=\ln \frac{\ln Q^{2}/\Lambda ^{2}}{\ln Q_{0}^{2}/\Lambda ^{2}}\;,
\end{equation}
where $Q_{0},\Lambda$ are scale parameters and we fixed it by
$Q_{0}^{2}=1 GeV^{2}$ and used $ \Lambda=0.203 GeV$. It is
completely known the moments of the unpolarized Singlet and
Non-singlet and for polarized ones we have
\begin{equation}
\Delta M^{NS}(n,Q^{2})=\exp (-\delta d_{NS}^{(0)n}s)\;,
\end{equation}
\[\Delta M^{S}(n,Q^{2})=\frac{1}{2}(1+\delta \rho )\exp (-\delta d_{+}s)\]
\begin{equation}
+\frac{1}{2}(1-\delta \rho )\exp (-\delta d_{-}s)\;,
\end{equation}
where $\delta \rho $ and other associated parameters are as
follows:
\begin{eqnarray}
\delta \rho &=&\frac{\delta d_{NS}^{(0)n}-\delta
d_{gg}^{(0)n}+4f\delta
d_{qg}^{(0)n}}{\Delta }\;,  \nonumber \\
\Delta &=&\delta d_{+}-\delta d_{-}=\sqrt{(\delta
d_{NS}^{(0)n}-\delta
d_{gg}^{(0)n})^{2}+8f\delta d_{qg}^{(0)n}\delta d_{gq}^{(0)n}}\;,  \nonumber \\
\delta d_{\pm } &=&\frac{1}{2}(\delta d_{NS}^{(0)n}+\delta
d_{gg}^{(0)n}\pm
\Delta )\;, \\
b &=&\frac{33-2f}{12\pi }\;,  \nonumber
\end{eqnarray}
$f$ is the number of active flavors. The anomalous dimensions $\delta d_{ij}^{(0)n}$  which are simply the $n$-th moment of polarized LO splitting function are given by
\cite{5}.
The moments of all polarized $u$ and $d$-valence quark in a proton are:
\begin{equation}
\Delta M_{u_{v}}(n,Q^{2})+\Delta M_{d_{v}}(n,Q^{2})=\Delta M^{NS}(n,Q^{2})\times
\Delta G_p^{NS}(n)
\end{equation}
The moment of polarized singlet distribution in a proton ($\Sigma $) is as follows:
\begin{equation}
\Delta M_{\Sigma }(n,Q^{2})=\Delta M^{S}(n,Q^{2})\times
\Delta G_p^{S}(n)\;.
\end{equation}
In Eq.(14) $\Sigma$ symbol indicates $\sum_{q=u,d,s}(q+\bar {q})$.
In this situation we can compute the moments of polarized sea quarks as:
\[
\Delta M_{\bar q }(n,Q^{2})=
\]
\begin{equation}
\frac{\Delta M_{\Sigma }(n,Q^{2})-\Delta M_{u_{v}}(n,Q^{2})-\Delta M_{d_{v}}(n,Q^{2})}{2f}\;.
\end{equation}
In moment space we can get the polarized moment of proton in terms
of their polarized constituent quarks as below
\be
{{\cal{M}}}(n,Q^2)= \frac{1}{2}\sum_{q=u,d,s}(e_q^2)(\Delta
M_{q}+ \Delta M_{\bar {q}}) \;.
\ee
By inserting Eqs. (13-15) in above equation, 16 unknown parameters
will be appeared which should be determined by a fitting processes.
\section{QCD fits to average of moments using Bernstein polynomials}
Because for a given value of $Q^2$, only a limited number of
experimental points, covering a partial range of values $x$, are
available, one can not use the moments equation like the present
one. A method device to deal to this situation is that to take
averages of structure functions with Bernstein
polynomials.\\
We define these polynomials as
\be p_{n,k}(x)=\frac{\Gamma (n+2) }{\Gamma (k+1) \Gamma (n+k+1)
}x^k(1-x)^{n-k}\;,\ee
 Thus we can compare theoretical predictions with experimental results for the Bernstein averages,
 which are defined by \cite{6}
 \be g_{n,k}(Q^2){\equiv}\int_{0}^{1}dxp_{nk}(x)g_1(x,Q^2)\;, \ee
In Eq. (17), $p_{n,k}(x) $ are normalized to unity,
$\int_{0}^{1}dxp_{n,k}(x)=1$.
 Using the binomial expansion in Eq.(17), it follows that the averages of $g_1$
   with $p_{n,k}(x)$ as weight functions, can be obtained in terms of odd and even moments,
\[
g_{n,k}=\frac{{(n-k)!}{\Gamma(n+2)}}{\Gamma(k+1)\Gamma(n-k+1)}\]
\be
\times \sum_{l=0}^{n-k}
\frac{(-1)^l}{l!(n-k-l)!}{{\cal{M}}({(k+l)+1}},Q^2)\;. \ee
where moments $\cal{M}$ are given by
\be
{{\cal{M}}({(k+l)+1}},Q^2)=\int_{0}^{1}x^{(k+l+1)-1}{g_1(x,Q^2)}dx\;,
 \ee
The integral (18)
represents an average of the function $g_{1}(x)$ in the region
${\bar{x}}_{n,k}-\frac{1}{2}\Delta{x}_{n,k}{\leq}x{\leq}{\bar{x}}_{n,k}+\frac{1}{2}\Delta{x}_{n,k}$
where ${\bar{x}}_{n,k}$ is the average of $x$ which is very near
to the maximum of $p_{n,k}(x)$, and $\Delta{x}_{n,k}$ is the
spread of ${\bar{x}}_{n,k}$.
 The key point is that
  values of $g_1$ outside this interval contribute little to the integral (18), as $p_{n,k}(x)$ decreases to zero very quickly. So, by suitably choosing
   $n$, $k$, we manage to adjust the region where the average is peaked to that in which we have experimental data. This means that we can use only 41 averages $g_{n,k}$ like
:
\begin{center}
$g_{2,1}^{(\exp )}(Q^{2}),g_{2,2}^{(\exp )}(Q^{2})$, ..., $g_{13,10}^{(\exp )}(Q^{2})$.
\end{center}
Other restriction which we assume here, is that to ignore the
effects of moments with high order $n$ which are not very
effective. To obtain these experimental averages from the E143 and
SMC data for $x{g_1}$ \cite{4}, we fit $x{g_1}(x,{Q^2})$ for each
bin in ${Q}^{2}$ separately, to the convenient phenomenological
expression \be
{xg_{1}}^{\hspace{-.12cm}{(phen)}}={\cal{A}}x^{\cal{B}}(1-x)^{\cal{C}}\;,
\ee this form ensures zero values for ${xg_{1}}$ at $x=0$, and
$x=1$. Using Eq.(21) with the fitted values of
${\cal{A}},{\cal{B}},{\cal{C}}$, one can then compute
${g}_{n,k}^{(exp)}({Q}^{2})$ using Eq.(18), in terms of Gamma
functions. Using Eq.(19) the Bernstein averages
${g}_{n,k}({Q}^{2})$ can be written in terms of odd and even
moments ${\cal{M}}(n,{Q}^{2})$, \ba
&&{g_{2,1}}(Q^2)=6\left({\cal{M}}(2,Q^2)-{\cal{M}}(3,Q^2)\right)\;,
\nonumber\\
&&{g_{2,2}}(Q^2)=3\left({\cal{M}}(3,Q^2)\right)\;,
\nonumber\\
&& \vdots\nonumber\\
&&{g_{13,10}}(Q^2)=4004.000001\left({\cal{M}}(11,Q^2)
-{\cal{M}}(14,Q^2)\right)\;,\nonumber\\
&&-12012\left({\cal{M}}(12,Q^2)-{\cal{M}}(13,Q^2)\right)\;. \ea We
shall use the  result of Eq.(21) for the QCD prediction of
${\cal{M}}(n,{Q}^{2})$. Thus there are 16 parameters to be simultaneously fitted to the
experimental ${g}_{n,k}({Q}^{2})$ averages. Defining a global
${\chi}^{2}$ for all the experimental data points, we
found an acceptable fit with minimum
 ${\chi}^{2}/{\rm{d.o.f.}}=0.03262/123$. Since polarized valon distribution are now determined with
16 $\it{known}$ parameters, the Eqs. (5,6) are now completely
determined for polarized NS and S valon distributions. In Fig.(1) we
plotted $y\times\delta G_p^{NS}$ and $y\times\delta G_p^{S}$ as a function of $y$.
\begin{figure}[tbh]
\centerline{\includegraphics[width=0.36\textwidth]{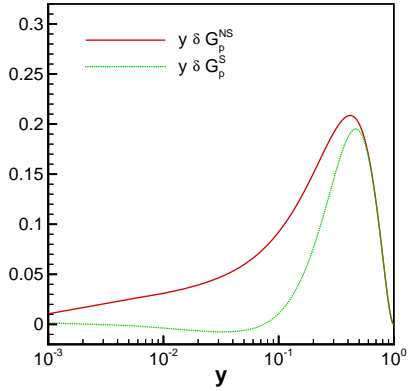}}
\caption{Non-singlet and Singlet polarized valon distributions as a
function of y.} \label{fig:fig1}
\end{figure}
\section{Polarized parton distributions and $\bf g_1^p$ structure function}
Now we want to compute the polarized structure function of proton. Since we calculated before polarized valon distribution in
a proton, by having the polarized structure function in a valon,
it is possible to extract polarized parton structure in a proton.
To obtain the $z$ dependence of  parton
distributions in  practical purposes from the $n-$%
dependent exact analytical solutions in Mellin-moment space, one
has to perform a numerical integral in order to invert the
Mellin-transformation. Consequently we can get the following expressions for
polarized parton distributions in a proton:

\ba \delta u_v(x,Q^2)&=&2\int_x^1  \delta
f^{NS}(\frac{x}{y},Q^2)\times\delta G_{U/p}(y)\frac{dy}{y}\;,\nonumber \\
\delta d_v(x,Q^2)&=&\int_x^1  \delta
f^{NS}(\frac{x}{y},Q^2)\times\delta G_{D/p}(y)\frac{dy}{y}\;,\nonumber \\
\delta \Sigma(x,Q^2)&=&\int_x^1  \delta
f^{S}(\frac{x}{y},Q^2)\times f(y)\;(2\delta G_{U/p}(y)\nonumber \\
&&+\delta G_{D/p}(y))\frac{dy}{y}\;, \ea

\begin{figure}[tbh]
\centerline{\includegraphics[width=0.35\textwidth]{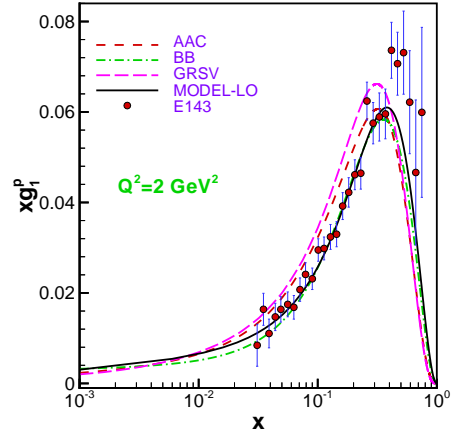}}
\caption{The results of $g_{1}^{p}(x,Q^{2})$, at $Q^2=2\; GeV^2$.
Our model is indicated by solid line. Other theoretical
predictions are from Ref.[3,7]. Experimental data points are from
Ref.[3]}
\end{figure}
 In leading order approximation in QCD, according to the quark model,
$g_{1}^{p}$ can be written as a linear combination of $\delta q$
and $\delta \overline{q}$,
\begin{equation}
g_{1}^{p}(x,Q^{2})=\frac{1}{2}\sum_{q} e_{q}^{2}[\delta
q(x,Q^{2})+\delta \overline{q}(x,Q^{2})]\;,
\end{equation}
where $e_{q}$ are the electric charges of the (light)
quark-flavors $q=u,d,s$.
We are now in a position to present the results for the proton
polarized structure function, $g_{1}^{p}$. We presented in Fig.(2)
the results of $g_{1}^{p}(x,Q^{2})$, for $Q^2=2\; GeV^2$ and compared it with
experimental data [4].
\section*{Acknowledgments}
We are grateful to R. C. Hwa for giving his usuful and
constructive comments. A.N.K thanks from Semnan university for
partial financial support to do this project. We acknowledge from
Institute for Studies in Theoretical Physics and Mathematics (IPM)
to support financially this project.

\end{document}